\documentclass[aip,jcp]{revtex4-1}

\usepackage{amsmath}
\usepackage{fullpage}
\usepackage{amssymb}
\usepackage{amsthm}
\usepackage{hyperref}
\usepackage{setspace}
\usepackage{graphicx}
\usepackage{bm}
\usepackage{physics}
\usepackage{enumerate}

\begin{document}


 \title{The chemical Langevin equation: a path integral view of Gillespie's derivation}

\author{John J. Vastola}
\email[john.j.vastola@vanderbilt.edu]{}
\affiliation{Department of Physics and Astronomy, Vanderbilt University, Nashville, Tennessee}
\affiliation{Quantitative Systems Biology Center, Vanderbilt University, Nashville, Tennessee}

\author{William R. Holmes}

\affiliation{Department of Physics and Astronomy, Vanderbilt University, Nashville, Tennessee}
\affiliation{Quantitative Systems Biology Center, Vanderbilt University, Nashville, Tennessee}
\affiliation{Department of Mathematics, Vanderbilt University, Nashville, Tennessee}

\date{\today}

\begin{abstract}
In 2000, Gillespie rehabilitated the chemical Langevin equation (CLE) by describing two conditions that must be satisfied for it yield a valid approximation of the chemical master equation (CME). In this work, we construct an original path integral description of the CME, and show how applying Gillespie's two conditions to it directly leads to a path integral equivalent to the CLE. We compare this approach to the path integral equivalent of a large system size derivation, and show that they are qualitatively different. In particular, both approaches involve converting many sums into many integrals, and the difference between the two methods is essentially the difference between using the Euler-Maclaurin formula and using Riemann sums. Our results shed light on how path integrals can be used to conceptualize coarse-graining biochemical systems, and are readily generalizable. 
\end{abstract}

\maketitle 



\section{Introduction}
\label{sec:intro}

Gillespie's classic paper \cite{gillespie2000} on how to derive the chemical Langevin equation (CLE) from the chemical master equation (CME) proceeds differently than by naively truncating the Kramers-Moyal expansion of the CME \cite{risken1987, luczka1995, risken1996} or by invoking the largeness of the system volume $\Omega$ \textit{a la} van Kampen \cite{vankampen1961, vankampen2007}; instead, he argues based on the existence of a \textit{time scale} with certain properties. In particular, his derivation completes avoids rewriting discrete number variables $n$ as concentration variables $x := n/\Omega$. 



By writing down two precise conditions that control the validity of the CLE (to be reviewed in Sec. \ref{sec:review}), he rehabilitated it as a well-founded approach to approximating stochastic dynamics described by the CME (in the face of ostensible no-go results like the Pawula theorem \cite{pawula1967}), and directly inspired the tau-leaping algorithm \cite{gillespie2001} and its many modifications \cite{gillespie2003, rathinam2003, cao2004, cao2005, cao2006, cao2007} for speeding up numerical simulations of biochemical reactions. 

Path integrals offer a way to think about stochastic processes that is somewhat independent from the usual differential equations perspective \cite{vastola2019}. This means that---at least in principle---there should be a way to translate Gillespie's derivation into path integral language. Because path integrals (along with associated technology like the renormalization group \cite{wilson1971pt1, wilson1971pt2, peskin1995, schwartz2014}) are known to be useful for understanding coarse-grained descriptions of systems (e.g. effective field theories\cite{georgi1994, schwartz2014}), such a translation should contribute meaningfully to our understanding of how to intelligently coarse-grain biochemical systems.  

In this paper, we show how Gillespie's two conditions translate to a path integral-based derivation of the chemical Langevin equation. Our approach here builds upon the path integral descriptions of Langevin/Fokker-Planck equations described in an earlier paper \cite{vastola2019}. We will proceed with little mathematical rigor (as is typical in physics), but with enough clarity that our arguments could in principle be made mathematically precise. 

The paper is organized as follows. In Sec. \ref{sec:review}, we review Gillespie's derivation of the CLE. In Sec. \ref{sec:pathint}, we construct a path integral description of CME dynamics. In Sec. \ref{sec:derivation}, we apply Gillespie's conditions to our path integral formulation to obtain the CLE, and also discuss an alternative method based on a large system volume argument. Finally, we discuss consequences of our work for understanding coarse-grained biochemical systems in Sec. \ref{sec:discussion}.


\section{Review of Gillespie's chemical Langevin equation derivation} \label{sec:review}

In this section, we review Gillespie's derivation \cite{gillespie2000} of the chemical Langevin equation from the chemical master equation.  We use the same notation Gillespie used in his paper, although we will not require that the same physical assumptions (i.e. well-stirred, dilute chemicals in a fixed volume and at constant temperature) hold, because the derivation does not depend on them.  

Consider a system with $N$ species and $M$ reactions. Denote the propensity function of the $j$th reaction by $a_j$, and the corresponding stoichiometry vector by $\boldsymbol{\nu}_j$. The chemical master equation reads
\begin{equation} \label{eq:cme}
\frac{\partial P(\mathbf{n}, t)}{\partial t} = \sum_{j=1}^M a_j(\mathbf{n} - \boldsymbol{\nu}_j) P(\mathbf{n} - \boldsymbol{\nu}_j, t) - a_j(\mathbf{n}) P(\mathbf{n}, t)
\end{equation}
where $P(\mathbf{n}, t)$ is the probability that the state of the system is $\mathbf{n} = (n_1, ..., n_N) \in \mathbb{N}^N$ at time $t$. 

Gillespie's derivation requires the existence of a time scale $\tau$ for which the following two conditions hold:
\begin{enumerate}[i]
\item The propensity functions do not change their values appreciably, \\i.e. $a_j(\mathbf{n}(t)) \approx a_j(\mathbf{n}(t'))$ for all $j$ and all $t' \in [t, t + \tau]$.
\item The average number of firings of each reaction over a time $\tau$ is much larger than $1$.
\end{enumerate}
Due to their connection with the tau-leaping algorithm \cite{gillespie2001, gillespie2003, rathinam2003, cao2004, cao2005, cao2006, cao2007} for approximately simulating CME dynamics, Gillespie later called these the first leap condition and the second leap condition \cite{gillespie2013}. They are in practice easily satisfied in the case of large molecule numbers, and they are exactly satisfied in the thermodynamic limit \cite{gillespie2009}, where the system volume $\Omega$ is taken to infinity while keeping all concentrations fixed. 

Consider $n_i(t)$, the number of molecules corresponding to species $i$ at time $t$. It changes in a small time $\Delta t$ according to
\begin{equation}
n_i(t + \Delta t) = n_i(t) + \sum_{j=1}^M \nu_{ji} K_j(a_j, \Delta t) \ ,
\end{equation}
where $\nu_{ji}$ is the $i$th component of the stoichiometry vector $\boldsymbol{\nu}_j$ (i.e. the change in number of species $i$ due to reaction $j$ firing once), and $K_j(a_j, \Delta t)$ is a random variable that describes the number of times reaction $j$ fires in $\Delta t$. 

For an arbitrary CME and arbitrary length of time $\Delta t$, $K_j$ might be taken from a complicated distribution. But if condition (i) holds in a length of time $\tau$, each reaction fires independently of each other reaction, because no reactions significantly change any propensity functions. Because (by definition) the probability of reaction $j$ firing in an infinitesimal time $dt$ is $a_j(n_i) dt$, and because that probability will not significantly change during the time length $\tau$, the number of times reaction $j$ fires in $\tau$ is well-approximated as a Poisson random variable with mean $a_j(n_i(t)) \tau$, which we will denote by $\mathcal{P}_j(a_j(n_i(t)) \tau)$. 

This means that when condition (i) holds we can write the time evolution of $n_i(t)$ over a length of time $\tau$ as
\begin{equation}
n(t + \tau) = n(t) + \sum_{j=1}^M \nu_{ji} \mathcal{P}_j(a_j(n_i(t)) \tau) \ .
\end{equation}
This equation is the basis for the tau-leaping approach first described by Gillespie in 2001 \cite{gillespie2001}, and later modified and extended by himself and others \cite{gillespie2003, rathinam2003, cao2004, cao2005, cao2006, cao2007}. 

If condition (ii) holds, the average number of times reaction $j$ fires in $\tau$ (i.e. $a_j(n_i(t)) \tau$) is much larger than $1$, so the Poisson random variables are well-approximated by normal random variables:
\begin{equation}
\mathcal{P}_j(a_j(n_i(t)) \tau) \approx \mathcal{N}_j(a_j(n_i(t)) \tau, a_j(n_i(t)) \tau) \ ,
\end{equation}
where $\mathcal{N}_j(a_j(n_i(t)) \tau, a_j(n_i(t)) \tau)$ is a normal random variable with mean and variance both equal to $a_j(n_i(t)) \tau$. If we also note that each normal random variable can be decomposed as
\begin{equation}
\mathcal{N}_j(a_j(n_i(t)) \tau, a_j(n_i(t)) \tau) = a_j(n_i(t)) \tau + \sqrt{a_j(n_i(t)) \tau} \ \mathcal{N}_j(0, 1) \ ,
\end{equation}
we can write the time evolution of $n_i(t)$ in a time $\tau$ as
\begin{equation}
n(t + \tau) = n(t) + \sum_{j=1}^M \nu_{ji} a_j(n_i(t)) \tau + \sum_{j=1}^M \nu_{ji} \sqrt{a_j(n_i(t)) \tau} \ \mathcal{N}_j(0, 1) \ .
\end{equation}
Because this equation has the form of an Euler-Maruyama time step, we can identify the dynamics of the system on the time scale $\tau$ with the set of $N$ stochastic differential equations (SDEs)
\begin{equation} \label{eq:CLE}
\dot{x}_i = \sum_{j=1}^M \nu_{ji} a_j(\mathbf{x})  + \sum_{j=1}^M \nu_{ji} \sqrt{a_j(\mathbf{x}) } \ \Gamma_j \ ,
\end{equation}
where the $\Gamma_j$ are $M$ independent Gaussian white noise terms, and where we have relabeled each $n_i$ as $x_i$ to emphasize that we are now working with continuous variables.  

Our chemical Langevin equation corresponds to a chemical Fokker-Planck equation
\begin{equation} \label{eq:CFP}
\frac{\partial P(\mathbf{x}, t)}{\partial t} = \sum_{i = 1}^N - \frac{\partial}{\partial x_i} \left[ \left( \sum_{j = 1}^M \nu_{ji} a_j(\mathbf{x}) \right) P(\mathbf{x}, t) \right]  + \frac{1}{2} \sum_{i = 1}^N  \sum_{i' = 1}^N   \frac{\partial^2}{\partial x_i \partial x_{i'}} \left[ \left( \sum_{j = 1}^M \nu_{ji} \nu_{ji'} a_j(\mathbf{x}) \right) P(\mathbf{x}, t) \right]
\end{equation}
which serves an approximation to the CME (Eq. \ref{eq:cme}). As Gillespie notes, this is exactly what one would get from truncating the Kramers-Moyal expansion of the CME at second order, so his derivation in some sense justifies the naive one. 

The CLE (Eq. \ref{eq:CLE}), and the associated chemical Fokker-Planck equation (Eq. \ref{eq:CFP}) describing how the system's probability density will evolve in time, are not without problems. They generically predict negative concentrations \cite{schnoerr2014} (although the hope is that the system has a negligibly small probability of occupying these states, and this is often borne out in practice), can be inaccurate for systems far from equilibrium \cite{grima2011}, may not always exhibit multistability when the CME is multistable \cite{duncan2015}, and can give rise to nonphysical probability currents at equilibrium \cite{ceccato2018}. 

Despite these shortcomings, utilizing the CLE can help speed up simulations of CME dynamics when some species have large molecule numbers \cite{winkelmann2017, harris2006, harris2009, iyengar2010} or when there is a clear separation of time scales \cite{haseltine2002, salis2005, wu2016}. Moreover, alternative schemes like the deterministic reaction rate equations and the linear noise approximation \cite{vankampen2007} can profitably be viewed as approximations to the CLE \cite{wallace2012}, and moment-closure approximations have comparable accuracy \cite{grima2012}. 

The CLE, and Langevin equations more generally, have become standard approaches to modeling noisy gene regulation \cite{wang2011, zhou2012, zhouli2016, simpson2004, hinczewski2016}. They have also been used to analyze noise-driven oscillations \cite{brett2014}, model intracellular calcium dynamics \cite{choi2010, rudiger2014, li2005}, study ion-channel gating \cite{huang2015}, and understand spiking neurons \cite{wallace2011}. While only approximate, the CLE is unquestionably useful.

\section{Path integral formulation of CME dynamics} \label{sec:pathint}

Although path integrals \cite{feynman1948} are most well-known in the context of quantum mechanics and quantum field theory \cite{feynman2010, schulman1981, kleinert2009, schwartz2014, shankar2011, shankar2017}, they have also proven useful for understanding classical stochastic phenomena like Brownian motion \cite{caldeira1983, lau2007, asheichyk2019}, conformational transitions \cite{wang2005, wang2006, orioli2017}, quantitative finance \cite{linetsky1997, ingber2000, kleinert2009}, population dynamics \cite{spanio2017, kamenev2008, tauber2011, tauber2012, serrao2017}, neuron firing \cite{ocker2017, iomin2019, baravalle2017, bressloff2015, bressloff2014hybrid, qiu2018, buice2013}, gene regulation \cite{wang2011, li2014, li2016landscape, li2013, wang2010, newby2014}, and chemical kinetics \cite{elgart2004, elgart2006, li2016twoscale, benitez2016, thomas2014}.

In this section, we will develop a straightforward path integral formulation of chemical master equation dynamics. Our path integral is constructed to closely resemble the formalism we used to describe SDE/Fokker-Planck dynamics in an earlier paper \cite{vastola2019}. To our knowledge, it is original, although certain aspects also resemble the approach used by Lazarescu et al. \cite{lazarescu2019}. The approach presented in this section is somewhat distinct from the often used Doi-Peliti approach \cite{doi1976, doi1976second, peliti1985, mattis1998}, which involves integrating over so-called coherent states and yields integrals instead of sums.

\subsection{States and operators}

Our main objective is to solve the CME, Eq. \ref{eq:cme}. Instead of solving it directly, we will solve a related problem phrased in terms of states and operators in a certain Hilbert space; this allows us to construct a path integral just as one does in quantum mechanics. 

Consider an infinite-dimensional Hilbert space spanned by the $\ket{\mathbf{n}}$ vectors (where $\mathbf{n} = (n_1, ..., n_N) \in \mathbb{N}^N$), in which an arbitrary state $\ket{\phi}$ is written
\begin{equation}
\ket{\phi} =  \sum_{n_1 = 0}^{\infty} \cdots \sum_{n_N = 0}^{\infty} c(\mathbf{n}) \ket{\mathbf{n}} 
\end{equation}
for some generally complex-valued coefficients $c(\mathbf{n})$. To ease notation, we will write
\begin{equation}
\sum_{\mathbf{n}} := \sum_{n_1 = 0}^{\infty} \cdots \sum_{n_N = 0}^{\infty}
\end{equation}
so that an arbitrary state reads
\begin{equation}
\ket{\phi} =  \sum_{\mathbf{n}}  c(\mathbf{n}) \ket{\mathbf{n}}  \ .
\end{equation}
Define an inner product in this space by
\begin{equation} \label{eq:innerprod_basis}
\braket{\mathbf{m}}{\mathbf{n}} = \delta_{\mathbf{m} \mathbf{n}}
\end{equation}
for all basis vectors $\ket{\mathbf{m}}$ and $\ket{\mathbf{n}}$, so that the inner product of two arbitrary states reads
\begin{equation} \label{eq:innerprod_arb}
\braket{\phi_2}{\phi_1} =  \sum_{\mathbf{n}} c_2^*(\mathbf{n}) c_1(\mathbf{n})  \ .
\end{equation}
Using the inner product defined by Eq. \ref{eq:innerprod_basis} and Eq. \ref{eq:innerprod_arb}, we can show that there is a resolution of the identity
\begin{equation}
1 = \sum_{\mathbf{n}} \ket{\mathbf{n}} \bra{\mathbf{n}}
\end{equation}
since $\braket{\mathbf{n}}{\phi} = c(\mathbf{n})$. Define the state operators $\hat{n}_i$ by 
\begin{equation}
\hat{n}_i \ket{\mathbf{n}} := n_i \ket{\mathbf{n}}
\end{equation}
for all $i = 1, ..., N$. We will associate any function $f(\mathbf{n}) = f(n_1, ..., n_N)$ with the operator $f(\hat{\mathbf{n}})$, whose action on a basis vector $\ket{\mathbf{n}}$ is
\begin{equation}
f(\hat{\mathbf{n}}) \ket{\mathbf{n}} := f(\hat{n}_1, ..., \hat{n}_N) \ket{\mathbf{n}}
\end{equation}
where there is no operator ordering ambiguity because the $\hat{n}_i$ all commute with one another. Also define the propensity function operators $\hat{a}_j$ via
\begin{equation}
\hat{a}_j \ket{\mathbf{n}} := a_j(\boldsymbol{n}) \ket{\mathbf{n} + \boldsymbol{\nu}_j}
\end{equation}
for all $j = 1, ..., M$, where $\boldsymbol{\nu}_j$ denotes the stoichiometry vector of the $j$th reaction. 

\subsection{Generating function and equation of motion}

In the spirit of Peliti\cite{peliti1985}, define the generating function
\begin{equation}
\ket{\psi(t)} := \sum_{\mathbf{n}}  P(\mathbf{n}, t) \ket{\mathbf{n}} 
\end{equation}
where, as in the previous section, $P(\mathbf{n}, t)$ is the probability that the state of the system is $\mathbf{n} = (n_1, ..., n_N)$ at time $t$. Note that
\begin{equation}
\begin{split}
\frac{\partial \ket{\psi}}{\partial t} &= \sum_{\mathbf{n}}  \frac{\partial P(\mathbf{n}, t)}{\partial t} \ket{\mathbf{n}}  \\
&= \sum_{\mathbf{n}}  \left[ \sum_{j=1}^M a_j(\mathbf{n} - \boldsymbol{\nu}_j) P(\mathbf{n} - \boldsymbol{\nu}_j, t) - a_j(\mathbf{n}) P(\mathbf{n}, t) \right] \ket{\mathbf{n}}  \\
&= \sum_{\mathbf{n}}   \sum_{j=1}^M a_j(\mathbf{n}) P(\mathbf{n}, t)  \ket{\mathbf{n}  + \boldsymbol{\nu}_j} -  \sum_{\mathbf{n}}   \sum_{j=1}^M a_j(\mathbf{n})  P(\mathbf{n}, t) \ket{\mathbf{n}}  \\
\end{split}
\end{equation}
where we have reindexed the left sum in the last step. Now we have
\begin{equation}
\begin{split}
\frac{\partial \ket{\psi}}{\partial t} &= \sum_{\mathbf{n}}  \left[ \sum_{j=1}^M a_j(\mathbf{n}) \ket{\mathbf{n}  + \boldsymbol{\nu}_j} - a_j(\mathbf{n})  \ket{\mathbf{n}} \right] P(\mathbf{n}, t) \\
&= \sum_{\mathbf{n}} \left[  \sum_{j=1}^M \hat{a}_j \ket{\mathbf{n}} - a_j(\mathbf{n}) \right] P(\mathbf{n}, t)  \ket{\mathbf{n}} \\
&= \left[  \sum_{j=1}^M \hat{a}_j - a_j(\hat{\mathbf{n}})  \right] \ket{\psi} \ .
\end{split}
\end{equation}
If we define the operator
\begin{equation} \label{eq:Hdef}
\hat{H} := \sum_{j=1}^M \hat{a}_j - a_j(\hat{\mathbf{n}}) \ ,
\end{equation}
which we will call (in analogy with quantum mechanics) the Hamiltonian, then we can write the equation describing the time evolution of the generating function as
\begin{equation} \label{eq:gfeom}
\frac{\partial \ket{\psi}}{\partial t} = \hat{H} \ket{\psi} \ .
\end{equation}
It is this equation that we will solve instead of the CME; since $\braket{\mathbf{n}}{\psi(t)} = P(\mathbf{n}, t)$, a solution to the CME can be extracted out of a solution to this equation.

\subsection{Deriving the CME path integral}

The formal solution to Eq. \ref{eq:gfeom} is
\begin{equation}
\ket{\psi(t_f)} = e^{\hat{H} (t_f - t_0)} \ket{\psi(t_0)} \ .
\end{equation}
At this point (following the usual procedure for deriving path integrals \cite{vastola2019}), we write the length of time $(t_f - t_0)$ as $T \Delta t$ for some large number of time steps $T$, split the propagator into many pieces, and insert many resolutions of the identity:
\begin{equation}
\begin{split}
\ket{\psi(t_f)} &= e^{\hat{H} \Delta t} \cdots e^{\hat{H} \Delta t} \ket{\psi(t_0)} \\
&= \sum_{\mathbf{n}_0}  \cdots  \sum_{\mathbf{n}_T} \ket{\mathbf{n}_T} \matrixel{\mathbf{n}_T}{e^{\hat{H} \Delta t}}{\mathbf{n}_{T-1}} \cdots \matrixel{\mathbf{n}_1}{e^{\hat{H} \Delta t}}{\mathbf{n}_0} \bra{\mathbf{n}_0} \ket{\psi(t_0)} \ .
\end{split}
\end{equation}
We are specifically interested in the transition probability $P(\mathbf{n}_f, t_f; \mathbf{n}_0, t_0)$. To obtain an expression for it, note that if $\ket{\psi(t_0)} = \ket{\mathbf{n}_0}$, then $P(\mathbf{n}_f, t_f; \mathbf{n}_0, t_0) = \braket{\mathbf{n}_f}{\psi(t_f}$. Hence, we have
\begin{equation} \label{eq:tprob_expansion}
P(\mathbf{n}_f, t_f; \mathbf{n}_0, t_0)  = \sum_{\mathbf{n}_1}  \cdots  \sum_{\mathbf{n}_{T-1}}  \matrixel{\mathbf{n}_T}{e^{\hat{H} \Delta t}}{\mathbf{n}_{T-1}} \cdots \matrixel{\mathbf{n}_1}{e^{\hat{H} \Delta t}}{\mathbf{n}_0}  \ .
\end{equation}
where $\mathbf{n}_T = \mathbf{n}_f$. Now we just need to evaluate these matrix elements and put them together. Choose $\Delta t$ sufficiently small so that
\begin{equation}\label{eq:smallHmel}
\begin{split}
\matrixel{\mathbf{n}_k}{e^{\hat{H} \Delta t}}{\mathbf{n}_{k-1}} &\approx \matrixel{\mathbf{n}_k}{1 + \hat{H} \Delta t}{\mathbf{n}_{k-1}} \\
&= \delta_{\mathbf{n}_k, \mathbf{n}_{k-1}} + \matrixel{\mathbf{n}_k}{\hat{H}}{\mathbf{n}_{k-1}} \Delta t  \ .
\end{split}
\end{equation}
We will take $\Delta t \to 0$ at the end of the calculation, so this equality will hold exactly. Using the specific form of $\hat{H}$ (Eq. \ref{eq:Hdef}), we have
\begin{equation}
\begin{split}
\matrixel{\mathbf{n}_k}{\hat{H}}{\mathbf{n}_{k-1}} &= \matrixel{\mathbf{n}_k}{\sum_{j=1}^M \hat{a}_j - a_j(\hat{\mathbf{n}}) }{\mathbf{n}_{k-1}} \\
&= \sum_{j=1}^M \matrixel{\mathbf{n}_k}{\hat{a}_j - a_j(\hat{\mathbf{n}}) }{\mathbf{n}_{k-1}} \\
&= \sum_{j=1}^M a_j(\mathbf{n}_{k-1}) \left[ \braket{\mathbf{n}_k}{\mathbf{n}_{k-1} + \boldsymbol{\nu}_j} - \braket{\mathbf{n}_k}{\mathbf{n}_{k-1}} \right] \\
&= \sum_{j=1}^M a_j(\mathbf{n}_{k-1}) \left[  \delta_{\mathbf{n}_k, \mathbf{n}_{k-1} + \boldsymbol{\nu}_j} - \delta_{\mathbf{n}_k, \mathbf{n}_{k-1}} \right] \ .
\end{split}
\end{equation}
Recall that the usual integral representation of the Dirac delta function reads
\begin{equation}
\delta_{\mathbf{m}, \mathbf{n}} = \int \frac{d\mathbf{p}}{(2\pi)^N} \ e^{- i \mathbf{p} \cdot (\mathbf{m} - \mathbf{n})} \ ,
\end{equation}
where $d\mathbf{p} = dp_1 \cdots dp_N$ and each $p_i$ is integrated over the whole real line. Using this representation, $\matrixel{\mathbf{n}_k}{\hat{H}}{\mathbf{n}_{k-1}}$ becomes
\begin{equation} \label{eq:Hmel_intform}
\int \frac{d\mathbf{p}_k}{(2\pi)^N} \ e^{- i \mathbf{p}_k \cdot (\mathbf{n}_k - \mathbf{n}_{k-1})} \left\{ \sum_{j=1}^M \left[ e^{i \mathbf{p}_k \cdot \boldsymbol{\nu}_j} - 1 \right] a_j(\mathbf{n}_{k-1}) \right\}
\end{equation}
where we have labeled the integration variable $\mathbf{p}_k$ to anticipate there being one integral for each matrix element in the final answer. Using Eq. \ref{eq:smallHmel}, $\matrixel{\mathbf{n}_k}{e^{\hat{H} \Delta t}}{\mathbf{n}_{k-1}}$ is approximately equal to
\begin{equation} \label{eq:Hmel_intform2}
\int \frac{d\mathbf{p}_k}{(2\pi)^N} \ e^{- i \mathbf{p}_k \cdot (\mathbf{n}_k - \mathbf{n}_{k-1})} \left\{ 1 + \Delta t \sum_{j=1}^M \left[ e^{i \mathbf{p}_k \cdot \boldsymbol{\nu}_j} - 1 \right] a_j(\mathbf{n}_{k-1}) \right\} \ .
\end{equation}
Noting that $\Delta t$ is small enough for the bracketed expression to be approximately equal to the corresponding exponential, our final expression for $\matrixel{\mathbf{n}_k}{e^{\hat{H} \Delta t}}{\mathbf{n}_{k-1}}$ becomes
\begin{equation} \label{eq:propmel_intform}
\int \frac{d\mathbf{p}_k}{(2\pi)^N} \ e^{- i \mathbf{p}_k \cdot (\mathbf{n}_k - \mathbf{n}_{k-1}) + \Delta t \sum_{j=1}^M \left[ \exp(i \mathbf{p}_k \cdot \boldsymbol{\nu}_j) - 1 \right] a_j(\mathbf{n}_{k-1})} \ .
\end{equation}
Using Eq. \ref{eq:tprob_expansion} and Eq. \ref{eq:propmel_intform},  we find that $P(\mathbf{n}_f, t_f; \mathbf{n}_0, t_0)$ can be written as the path integral
\begin{equation} \label{eq:pathint_CME}
P = \lim_{T \to \infty} \sum_{\mathbf{n}_1} \cdots \sum_{\mathbf{n}_{T-1}} \int  \frac{d\mathbf{p}_1}{(2\pi)^N} \cdots \int \frac{d\mathbf{p}_T}{(2\pi)^N} \ \exp{\sum_{k = 1}^{T} - i \mathbf{p}_k \cdot (\mathbf{n}_k - \mathbf{n}_{k-1}) + \Delta t \sum_{j=1}^M \left[ e^{i \mathbf{p}_k \cdot \boldsymbol{\nu}_j} - 1 \right] a_j(\mathbf{n}_{k-1})} 
\end{equation}
which resembles the MSRJD (Martin-Siggia-Rose-De Dominicis) path integral description \cite{msr1973, janssen1976, dd1976, ddpeliti1978, hertz2016} of the Fokker-Planck equation. Again, while the Doi-Peliti path integral involves integrating over coherent states, this path integral involves integrating over every possible discrete path through $\mathbb{N}^N$ that goes from $\mathbf{n}_0$ to $\mathbf{n}_f$. 

\section{Path integral derivation of the chemical Langevin equation}
\label{sec:derivation}

In this section, we reinterpret Gillespie's derivation of the CLE in the context of stochastic path integrals, and show explicitly how his two conditions translate in the path integral context. Our central tool will be the Euler-Maclaurin formula \cite{abramowitz1974, lampret2001}, which allows one to approximate sums as integrals (plus correction terms). It says that
\begin{equation} \label{eq:eulermaclaurin}
\sum_{n=a}^b f(n) \sim \int_a^b f(x)\,dx + \frac{f(b) + f(a)}{2} + \sum_{k=1}^\infty \,\frac{B_{2k}}{(2k)!} \left[ f^{(2k - 1)}(b) - f^{(2k - 1)}(a) \right]
\end{equation}
where $B_{2k}$ is the $(2k)$th Bernoulli number, and the ``$\sim$'' symbol is to indicate that we are to interpret the right-hand side as an asymptotic expansion (generically, the infinite sum may not be convergent, but retaining a finite number of terms still usually provides a good approximation to the left-hand side).

The Euler-Maclaurin formula is not an unfamiliar tool in chemical physics, given that it is often used to approximate partition functions \cite{pathria2011, kayser1978, liao2015} to good accuracy in certain regimes (e.g. the high temperature limit). It has also been used for other interesting purposes, like computing Fermi-Dirac integrals \cite{rzadkowski2008}, and proving the asymptotic equivalence of two descriptions of Coulombic systems in certain potentials \cite{cioslowski2013}.  

Roughly speaking, we will proceed as follows. Condition (i) will allow us to approximate each sum in Eq. \ref{eq:pathint_CME} as an integral, and to argue that the correction terms are small; meanwhile, condition (ii) will allow us to Taylor expand the $\exp(i \mathbf{p}_k \cdot \boldsymbol{\nu}_j)$ terms in Eq. \ref{eq:pathint_CME} to second order in the momenta $\mathbf{p}_k$. The result of these two approximations will be a MSRJD path integral, which we know from studies of stochastic path integrals \cite{vastola2019} to be equivalent to a system of Langevin equations. In particular, it will be equivalent to Eq. \ref{eq:CLE}, the CLE. 

\subsection{Only some paths satisfy Gillespie's conditions}
\label{sec:dpaths}

Gillespie's first condition (see Sec. \ref{sec:review}) says that, in a period of time $\tau$, the propensity functions do not change appreciably. Upon some reflection, we realize that this cannot be true for \textit{all} possible trajectories the system might have, assuming the propensity functions have some state-dependence (which, in general, they do). In principle, it is possible that the number of molecules of some species jumps between $1$ and $10^{100}$, wildly and irregularly, so that there does not exist any time scale on which the propensity functions do not change appreciably. Indeed, all sorts of crazy trajectories are possible \textit{in principle}---but they are overwhelmingly unlikely in practice. 

While there certainly exist crazy and pathological paths for which it is hard or impossible to find a time scale $\tau$ that satisfies Gillespie's first condition, the requirement is not so stringent for most of the trajectories the system might take. In other words, we will suppose that the first condition is satisfied for the \textit{dominant paths} rather than for all paths. 

A similar argument applies to the second condition. This means that, in applying our two conditions, we will no longer be summing over all possible paths (c.f. Eq. \ref{eq:pathint_CME}). Instead, we will be summing over all possible paths that satisfy the two conditions, a collection which we will assume includes the dominant or most likely paths. 

If we are not summing over \textit{all} possible paths, what does our region of integration look like? To understand this, it is helpful to consider the simple case of a CME with one species and one reaction. Label the number of that species by $n$, and the propensity function of the single reaction by $a$. 

Imagine starting the system in the state with $n_0$ molecules and thinking about where it will go (i.e. all possible states $n_1$) in the next time length $\tau$. For the dominant paths, we assume that the difference $|a(n_1) - a(n_0)|$ is small, so that the propensity function did not change appreciably. But what do we mean by ``appreciably''? 

In a paper showing his two conditions hold in the thermodynamic limit \cite{gillespie2009}, Gillespie assumed that his first condition meant
\begin{equation} \label{eq:gill_app}
\frac{|a(n_1) - a(n_0)|}{a(n_0)} \ll 1
\end{equation}
i.e. that the change in the propensity function on the time scale $\tau$ is negligible compared to the size of its original value. This matches the intuition we have about what constitutes a negligible change in population size: for example, if the population size changed by $100$ molecules, but the total number of molecules is on the order of $10^5$, we imagine that change not to be noticeable. 

Here, we can be a little bit more precise than Eq. \ref{eq:gill_app}. We generally assume that our propensity functions are nicely behaved---in particular, that they are continuous, that they are infinitely differentiable, and that we may freely Taylor expand them. That is, we assume the $a_j$ are analytic functions throughout our domain. Because most propensity functions of interest are polynomials (or at worst, rational functions like Hill functions), and because expressions like the Kramers-Moyal expansion already assume the $a_j$ are smooth, these assumptions do not turn out to be particularly strong. 

Suppose $a(n_0) > 0$, which is always true in the regime we care about, since we will usually need $n$ sufficiently large; generic monomolecular and bimolecular propensity functions have zeros at $n = 0$ and $n = 1$. Because $a$ is continuous, for any $\epsilon > 0$ we can find a $\delta > 0$ such that
\begin{equation} \label{eq:continuity1}
|a(n_1) - a(n_0)| < \epsilon a(n_0)
\end{equation}
provided $| n_1 - n_0 | < \delta$. In order for the correction terms that arise from applying the Euler-Maclaurin formula (Eq. \ref{eq:eulermaclaurin}) to be negligible, we also want to bound the derivatives of $a$ in a similar fashion.

Analogous conditions apply in the general case, where the $a_j$ may be functions of more than one variable. The moral of the story is that, because of the assumed smooth behavior of the propensity functions, we can find a region where they (and their derivatives) do not vary appreciably. In the simple one-dimensional case, this is an `interval' $[n_0 - \delta_0^-, n_0 + \delta_0^+] \subseteq \mathbb{N}$ (where we let $\delta_0^- \neq \delta_0^+$ in general since we need $n_0 - \delta_0^-$ and $n_0 + \delta_0^+$ to both be natural numbers); in general, this is the intersection of an open set with a lattice: $U_0(\delta) \cap \mathbb{N}^N \subseteq \mathbb{N}^N$. For convenience, we will use $U_0$ to denote both the open set and its lattice intersection. 

Hence, for a one-dimensional system, we restrict ourselves to paths
\begin{equation} \label{eq:1D_path_restrict}
\sum_{n_1 = 0}^{\infty} \cdots \sum_{n_{N-1} = 0}^{\infty}  \to \sum_{n_1 = n_0 - \delta_0^-}^{n_0 + \delta_0^+} \cdots \sum_{n_{N-1} = n_{N-2} - \delta_{N-2}^-}^{n_{N-2} + \delta_{N-2}^+} 
\end{equation}
where the $\delta_i^+$ and $\delta_i^-$, as in the discussion above, are chosen so that the propensity functions and their derivatives vary within acceptable bounds. For an arbitrary CME, we restrict ourselves to paths
\begin{equation} \label{eq:ND_path_restrict}
\sum_{\mathbf{n}_1} \cdots \sum_{\mathbf{n}_{T-1}} \to \sum_{\mathbf{n}_1 \in U_0} \cdots \sum_{\mathbf{n}_{T-1} \in U_{T-2}}
\end{equation}
where the sets $U_i \subseteq \mathbb{N}^N$ are chosen similarly.

 \subsection{Coarse-graining time}
\label{sec:cgtime}

There is another `philosophical' point we need to address. Earlier, we imagined breaking up the propagator into $T$ time steps of length $\Delta t$, and choosing $T$ to be large enough (or equivalently, $\Delta t$ to be small enough) that each piece of the propagator was well-approximated by its first-order Taylor expansion (c.f. Eq. \ref{eq:smallHmel}). However, Gillespie's two conditions only apply on the `coarser' time scale $\tau$. How do we go from time steps of size $\Delta t$ to time steps of size $\tau$ in Eq. \ref{eq:pathint_CME}?

There are two straightforward ways we can imagine. The simpler way is to say that, since we are in the business of making approximations \textit{anyway}, we may as well make the approximation that Eq. \ref{eq:pathint_CME} is valid on the time scale $\tau$, and that the terms we neglected when Taylor expanding the propagator do not matter much in the regime where Gillespie's conditions apply. 

But there is a more intellectually honest way to proceed. Suppose we originally broke the propagator into $S \cdot T$ time steps, for some natural number $S$ large enough for our derivation to go through without issue. This means that the time step in our path integral has size $\Delta t := t/(S \cdot T)$. We want to rewrite our path integral in terms of a `macroscopic' time scale $\tau := t/T$, which corresponds to breaking up the overall time $t$ into $T$ time steps of length $\tau$. 

Schematically, this means we want to make the following identifications:
\begin{equation}
\begin{split}
\mathbf{n}_0 \xrightarrow{\Delta t} \mathbf{n}_1 \xrightarrow{\Delta t} \cdots \xrightarrow{\Delta t} \mathbf{n}_S \ &: \ \mathbf{n}_0 \xrightarrow{\tau} \mathbf{n}_1 \\
\mathbf{n}_S \xrightarrow{\Delta t} \mathbf{n}_{S+1} \xrightarrow{\Delta t} \cdots \xrightarrow{\Delta t} \mathbf{n}_{2S} \ &: \ \mathbf{n}_1 \xrightarrow{\tau} \mathbf{n}_2 \\
&\vdots \\
\mathbf{n}_{S \cdot (T-1)} \xrightarrow{\Delta t} \mathbf{n}_{S \cdot (T-1)+1} \xrightarrow{\Delta t} \cdots \xrightarrow{\Delta t} \mathbf{n}_{S \cdot T} \ &: \ \mathbf{n}_{T-1} \xrightarrow{\tau} \mathbf{n}_{T}
\end{split}
\end{equation}
The argument of the exponential in Eq. \ref{eq:pathint_CME} reads
\begin{equation} \label{eq:exparg}
\sum_{k = 1}^{S \cdot T} - i \mathbf{p}_k \cdot (\mathbf{n}_k - \mathbf{n}_{k-1}) + \Delta t \sum_{j=1}^M \left[ e^{i \mathbf{p}_k \cdot \boldsymbol{\nu}_j} - 1 \right] a_j(\mathbf{n}_{k-1}) \ .
\end{equation}
Consider the following small piece of this expression:
\begin{equation} \label{eq:smallpiece}
\sum_{k = 1}^{S \cdot T} \left[ e^{i \mathbf{p}_k \cdot \boldsymbol{\nu}_j} - 1 \right] a_j(\mathbf{n}_{k-1}) \ .
\end{equation}
Assuming (on the dominant paths) that the propensity function $a_j$ only changes appreciably on the time scale $\tau = S \Delta t$, we can make the approximation that
\begin{equation}
\begin{split}
a_j(\mathbf{n}_0) \approx a_j(\mathbf{n}_1) \approx  &\cdots \approx a_j(\mathbf{n}_{S-1}) \\
a_j(\mathbf{n}_S) \approx a_j(\mathbf{n}_{S+1}) \approx  &\cdots \approx a_j(\mathbf{n}_{2S - 1}) \\
&\vdots \\
a_j(\mathbf{n}_{S \cdot (T-1)}) \approx a_j(\mathbf{n}_{S \cdot (T-1) + 1}) \approx  &\cdots \approx a_j(\mathbf{n}_{S \cdot T - 1}) 
\end{split}
\end{equation}
and rewrite Eq. \ref{eq:smallpiece} in terms of $a_j(\mathbf{n}_0), a_j(\mathbf{n}_S), a_j(\mathbf{n}_{2S}), ..., a_j(\mathbf{n}_{S \cdot T})$ only. This means that the only places the `intermediate' time steps (e.g. $\mathbf{n}_1, ..., \mathbf{n}_{S-1}$, or $\mathbf{n}_{S+1}, ..., \mathbf{n}_{2S-1}$) will appear are in the piece that reads
\begin{equation} \label{eq:smallpiece2}
\sum_{k = 1}^{S \cdot T} - i \mathbf{p}_k \cdot (\mathbf{n}_k - \mathbf{n}_{k-1}) \ .
\end{equation}
Happily, this means that all of the intermediate time steps can be summed over. For example, 
\begin{equation}
\sum_{\mathbf{n}_1} \cdots \sum_{\mathbf{n}_{S-1}} \exp\left\{ \sum_{k = 1}^{S} - i \mathbf{p}_k \cdot (\mathbf{n}_k - \mathbf{n}_{k-1}) \right\} \approx \delta(\mathbf{p}_1 - \mathbf{p}_2) \delta(\mathbf{p}_2 - \mathbf{p}_3)  \cdots \delta(\mathbf{p}_{S-1} - \mathbf{p}_S) 
\end{equation}
where the right-hand side is approximate because, due to our restriction of the sum domain in the previous section, the sum representation of the Dirac delta function
\begin{equation}
\frac{1}{(2\pi)^N} \sum_{\mathbf{n}} \exp\left\{ - i \mathbf{n} \cdot (\mathbf{p} - \mathbf{p}') \right\} = \delta(\mathbf{p} - \mathbf{p}')
\end{equation}
only approximately applies. After summing over all intermediate time steps and integrating out extraneous $\mathbf{p}_k$ using the delta functions that appear, Eq. \ref{eq:smallpiece} reads
\begin{equation} \label{eq:exparg2}
\begin{split}
& \sum_{k = 1}^{T} - i \mathbf{p}_k \cdot (\mathbf{n}_k - \mathbf{n}_{k-1}) + S \Delta t \sum_{j=1}^M \left[ e^{i \mathbf{p}_k \cdot \boldsymbol{\nu}_j} - 1 \right] a_j(\mathbf{n}_{k-1}) \\
= & \sum_{k = 1}^{T} - i \mathbf{p}_k \cdot (\mathbf{n}_k - \mathbf{n}_{k-1}) + \tau \sum_{j=1}^M \left[ e^{i \mathbf{p}_k \cdot \boldsymbol{\nu}_j} - 1 \right] a_j(\mathbf{n}_{k-1}) \ .
\end{split}
\end{equation}
Hence, using Gillespie's first condition, we have successfully gone from a path integral with time scale $\Delta t$ to a path integral with a `coarser' time scale $\tau$.

\subsection{Applying condition 1}

In this section, we will apply condition (i) in order to convert the sums in Eq. \ref{eq:pathint_CME} to integrals. After restricting our domain to the dominant paths (see Sec. \ref{sec:dpaths}) and coarse-graining time (see Sec. \ref{sec:cgtime}), our current path integral description of the CME reads
\begin{equation} \label{eq:pathint_CME_tau}
P \approx \sum_{\mathbf{n}_1 \in U_0} \cdots \sum_{\mathbf{n}_{T-1} \in U_{T-2}} \int  \frac{d\mathbf{p}_1}{(2\pi)^N} \cdots \int \frac{d\mathbf{p}_T}{(2\pi)^N} \ \exp{ - S \tau} 
\end{equation}
where we recall that the sets $U_0, ..., U_{T-2}$ cover all trajectories on which Gillespie's two conditions apply, and where we have defined the function (which we can call the ``action'', in analogy with quantum mechanics)
\begin{equation} \label{eq:action}
S := \sum_{k = 1}^{T} i \mathbf{p}_k \cdot \left( \frac{\mathbf{n}_k - \mathbf{n}_{k-1}}{\tau} \right) - \sum_{j=1}^M \left[ e^{i \mathbf{p}_k \cdot \boldsymbol{\nu}_j} - 1 \right] a_j(\mathbf{n}_{k-1})
\end{equation}
to ease notation. We will proceed using the Euler-Maclaurin formula (Eq. \ref{eq:eulermaclaurin}). As a starting point, consider Eq. \ref{eq:pathint_CME_tau} in one-dimension:
\begin{equation} \label{eq:pathint_CME_tau1D}
P \approx \sum_{n_1 = n_0 - \delta_0^-}^{n_0 + \delta_0^+} \cdots \sum_{n_{N-1} = n_{N-2} - \delta_{N-2}^-}^{n_{N-2} + \delta_{N-2}^+}  \int  \frac{dp_1}{2\pi} \cdots \int \frac{dp_T}{2\pi} \ \exp{ - S \tau} 
\end{equation}
where the $\delta_i^-$ and $\delta_i^+$ are as described in Sec. \ref{sec:dpaths}. Using the Euler-Maclaurin formula, we have
\begin{equation}
\begin{split}
& \sum_{n_1 = n_0 - \delta_0^-}^{n_0 + \delta_0^+} \exp{ - S \tau} \\
\approx & \int_{n_0 - \delta_0^-}^{n_0 + \delta_0^+} \exp{ - S \tau} \ dn_1 + \frac{e^{ - S(n_0 + \delta_0^+) \tau} + e^{ - S(n_0 - \delta_0^-) \tau}}{2} + \sum_{k=1}^\infty \frac{B_{2k}}{(2k)!} \frac{d^{2k-1}}{dn_1^{2k-1}} \left[ e^{ - S \tau} \right]_{n_0 - \delta_0^-}^{n_0 + \delta_0^+} \ .
\end{split}
\end{equation}
Now we need to argue that the correction terms can safely be neglected. Define $\delta := \delta^+_0 + \delta^-_0$. Because the propensity functions don't change vary much in the interval $[n_0 - \delta^-_0, n_0 + \delta^+_0]$ (by Gillespie's first condition), the integral term is roughly
\begin{equation}
\exp\left\{ - S(n_0) \tau \right\} \delta \ .
\end{equation}
Meanwhile, the next term is roughly
\begin{equation}
\exp\left\{ - S(n_0) \tau \right\} 
\end{equation}
which should be negligible compared to the first as long as $\delta \gg 1$. This should certainly be true; if $\delta \sim 1$, our conditions are either too strict, or we are in a regime with too small molecule numbers. 

Because the propensity functions $a_j$ do not change much (and because the $a_j$ are nicely behaved, usually monotonic functions in the regime we care about), they are approximately `flat'. This means that their derivatives $a^{(2k-1)}_j$ are small. For example,
\begin{equation}
\begin{split}
\frac{d}{dn_1} \left[ e^{- S \tau} \right] &= \left[ i(p_2 - p_1) + \tau \sum_{j=1}^M \left[ e^{i p_k \nu_j} - 1 \right] a_j' \right] e^{- S \tau}
\end{split}
\end{equation}
so
\begin{equation}
\begin{split}
& \frac{d}{dn_1} \left[ e^{- S \tau} \right]_{n_0 - \delta_0^-}^{n_0 + \delta_0^+} \\ 
=& \tau  \sum_{j=1}^M \left[ e^{i p_k \nu_j} - 1  \right] \left[ a_j'(n_0 + \delta_0^+) e^{- S(n_0 + \delta_0^+) \tau} - a_j'(n_0 - \delta_0^-) e^{- S(n_0 - \delta_0^-) \tau} \right]   \\
 \approx & \tau e^{- S(n_0) \tau} \sum_{j=1}^M \left[ e^{i p_k \nu_j} - 1  \right] \left[ a_j'(n_0 + \delta_0^+) - a_j'(n_0 - \delta_0^-) \right]   \\
 \approx & 0 \ .
\end{split}
\end{equation}
In summary, we have
\begin{equation}
\begin{split}
\sum_{n_1 = n_0 - \delta_0^-}^{n_0 + \delta_0^+} \exp{ - S \tau} \approx  \int_{n_0 - \delta_0^-}^{n_0 + \delta_0^+} \exp{ - S \tau} \ dn_1 
\end{split}
\end{equation}
which means we've successfully converted a sum into an integral. Apply this argument many more times to obtain
\begin{equation}
\begin{split}
& \sum_{n_1 = n_0 - \delta_0^-}^{n_0 + \delta_0^+} \cdots \sum_{n_{T-1} = n_{T-2} - \delta_{T-2}^-}^{n_{T-2} + \delta_{T-2}^+} \exp{ - S \tau} \\
\approx & \int_{n_0 - \delta_0^-}^{n_0 + \delta_0^+} dn_1 \cdots \int_{n_{T-2} - \delta_{T-2}^-}^{n_{T-2} + \delta_{T-2}^+} dn_{T-1} \  \exp{ - S \tau} \ .
\end{split}
\end{equation}
A similar argument applies to the $N$ species path integral (Eq. \ref{eq:pathint_CME_tau}); the only difference is that the Euler-Maclaurin formula must be applied $N$ times for each time step, because we would like to convert $N$ sums to an $N$-variable integral. 

Alternatively, one can argue using the appropriate many sum generalization of the Euler-Maclaurin formula (Eq. \ref{eq:eulermaclaurin}). There is some literature on generalizations of it to sums over polytopes\cite{karshon2003, guillemin2007, tate2011} (schematically, shapes in $N$-dimensional space whose vertices we can imagine as living in $\mathbb{Z}^N$). The main challenge for this approach would be to show that satisfying Gillespie's first condition corresponds to satisfying the requirements associated with the approximation being accurate (which are somewhat more technical than those for the single sum Euler-Maclaurin formula). 

The end result of all this is
\begin{equation} \label{eq:pathint_CME_tau_ints}
P \approx \int_{U_0} d\mathbf{x}_1 \cdots \int_{U_{T-2}} d\mathbf{x}_{T-1} \int  \frac{d\mathbf{p}_1}{(2\pi)^N} \cdots \int \frac{d\mathbf{p}_T}{(2\pi)^N} \ \exp{ - S \tau} \ .
\end{equation}
where we have relabeled each $\mathbf{n}_k$ as $\mathbf{x}_k$ to (as in Sec. \ref{sec:review}) emphasize that we are now working with continuous variables. We remark that, if not for the bounds, we would have a Kramers-Moyal path integral (see Sec. V of our earlier paper \cite{vastola2019}).

\subsection{Applying condition 2}
\label{sec:con2}

Consider the terms in the action $S$ (Eq. \ref{eq:action}) that look like
\begin{equation}
\left[ e^{i \mathbf{p}_k \cdot \boldsymbol{\nu}_j} - 1 \right] a_j(\mathbf{x}_{k-1}) \tau \ .
\end{equation}
Condition (ii) tells us that, for the dominant paths, $a_j(\mathbf{x}_{k-1}) \tau \gg 1$. In particular, we will assume that it is \textit{so} large that Taylor expanding the term it is multiplied by will have a negligible effect on the overall value, i.e.
\begin{equation}
\begin{split}
& \left[ e^{i \mathbf{p}_k \cdot \boldsymbol{\nu}_j} - 1 \right] a_j(\mathbf{x}_{k-1}) \tau  \\
 \approx & \left[ i \mathbf{p}_k \cdot \boldsymbol{\nu}_j - \frac{1}{2} \sum_{\ell = 1}^N \sum_{\ell' = 1}^N  p_k^{\ell} p_k^{\ell'} \nu_{j \ell} \nu_{j \ell'}  \right] a_j(\mathbf{x}_{k-1}) \tau   \ .
 \end{split}
\end{equation}
where $p_k^{\ell}$ is the $\ell$-th component of the vector $\mathbf{p}_k$. Thus, we finally obtain
\begin{equation} \label{eq:MSRJD_restricted}
S \approx \sum_{k = 1}^{T} i \mathbf{p}_k \cdot \left[ \frac{\mathbf{x}_k - \mathbf{x}_{k-1}}{\tau} - \sum_{j=1}^M \boldsymbol{\nu}_j a_j(\mathbf{x}_{k-1}) \right] + \frac{1}{2} \sum_{\ell = 1}^N \sum_{\ell' = 1}^N \sum_{j=1}^M p_k^{\ell} p_k^{\ell'} \nu_{j \ell} \nu_{j \ell'} a_j(\mathbf{x}_{k-1})
\end{equation}
which looks just like the action for the MSRJD path integral (see Sec. V of our earlier paper\cite{vastola2019}) corresponding to the chemical Fokker-Planck equation (Eq. \ref{eq:CFP}). Our final result for the whole path integral reads
\begin{equation} \label{eq:pathint_CME_final}
\begin{split}
P \approx & \int_{U_0} d\mathbf{x}_1 \cdots \int_{U_{T-2}} d\mathbf{x}_{T-1} \int  \frac{d\mathbf{p}_1}{(2\pi)^N} \cdots \int \frac{d\mathbf{p}_T}{(2\pi)^N} \ \\
& \exp{- \sum_{k = 1}^{T} \left[ i  \mathbf{p}_k \cdot \left( \frac{\mathbf{x}_k - \mathbf{x}_{k-1}}{\tau} - \sum_{j=1}^M \boldsymbol{\nu}_j a_j(\mathbf{x}_{k-1}) \right) +  \frac{1}{2} \sum_{\ell = 1}^N \sum_{\ell' = 1}^N \sum_{j=1}^M p_k^{\ell} p_k^{\ell'} \nu_{j \ell} \nu_{j \ell'} a_j(\mathbf{x}_{k-1}) \right] \tau}
\end{split}
\end{equation}
which looks like the usual Fokker-Planck path integral but with restricted integration bounds. 

The appearance of Eq. \ref{eq:pathint_CME_final} can be compacted somewhat if we define the diffusion tensor $D_{\ell \ell'}$:
\begin{equation} \label{eq:dtensor}
D_{\ell \ell'}(\mathbf{x}) := \frac{1}{2} \sum_{j=1}^M \nu_{j \ell} \nu_{j \ell'} a_j(\mathbf{x}) \ .
\end{equation}
At the CLE/Fokker-Planck level, the diffusion tensor captures all information about a system's noise. It must be positive semidefinite for the Fokker-Planck equation and its corresponding path integral to make sense \cite{gardiner2009, risken_fokker-planck_1996}. 

Finish the derivation by enlarging our integration domain as much as possible (while keeping the diffusion tensor positive semidefinite), assuming that permitting these additional paths does not substantially contribute to transition probabilities, since they were small enough to neglect in the first place. In general, we do not expect that the appropriate domain for our new continuous variables will be $[0, \infty)^N$, despite the fact that our original domain was $\mathbb{N}^N$. For example, the chemical Langevin equation \cite{gillespie2000} for the birth-death process (with birth rate $k$, death rate $\gamma$, and steady state mean $\mu := k/\gamma$) reads
\begin{equation}
\dot{x} = k - \gamma x + \sqrt{k + \gamma x} \ \eta(t)
\end{equation}
and is naturally defined on $[-\mu, \infty)$, because there is always some nonzero probability that the noise term will push the system into negative concentrations while its magnitude is greater than or equal to zero, i.e. when $k + \gamma x = \gamma (\mu + x) \geq 0$.

\subsection{Comparison with the system volume approach}
\label{sec:volume}

We have shown in the previous few sections how Gillespie's derivation works in a path integral context. Because Gillespie himself \cite{gillespie2000} compared his approach to ones which rely upon the largeness of the system volume $\Omega$, it is interesting to do that here also. Let us translate the typical system volume approach into path integral language, and see how it compares with the approach we described earlier. 

Consider again a CME with $N$ species and $M$ reactions (Eq. \ref{eq:cme}), but this time with the additional physical context that the chemicals interact inside a very large volume $\Omega$. Suppose we rewrite the CME in terms of concentration variables $x_i := n_i/\Omega$ for all $i = 1, ..., N$. The change in variables will lead to the probability density function $P(\mathbf{n}, t)$ increasing by a factor of $\Omega^N$: 
\begin{equation} \label{eq:Pchange}
\begin{split}
P(\mathbf{n}, t) d\mathbf{n} &= \Omega^N P(\mathbf{n}, t) d\mathbf{x} = P(\mathbf{x}, t) d\mathbf{x}  \\
\implies \ P(\mathbf{x}, t) &= \Omega^N P(\mathbf{n}, t)  \ .
\end{split}
\end{equation}
Gillespie used rigorous microphysical arguments \cite{gillespie1976, gillespie1992, gillespiebook1992, gillespie2000} to show that the volume-dependence of the propensity functions for monomolecular, bimolecular, and trimolecular reactions goes like
\begin{equation} \label{eq:propV}
a_j(\mathbf{n}) = \Omega \ \tilde{a}_j(\mathbf{x})
\end{equation}
where the adjusted propensity functions $\tilde{a}_j$ are volume-independent. Using Eq. \ref{eq:Pchange} and Eq. \ref{eq:propV}, our original CME path integral (Eq. \ref{eq:pathint_CME}) can be rewritten as
\begin{equation} \label{eq:pathint_CME_V}
\begin{split}
P = &\lim_{T \to \infty} \Omega^N \sum_{\mathbf{n}_1} \cdots \sum_{\mathbf{n}_{T-1}} \int  \frac{d\mathbf{p}_1}{(2\pi)^N} \cdots \int \frac{d\mathbf{p}_T}{(2\pi)^N} \\
& \exp{\sum_{k = 1}^{T} - i \Omega \ \mathbf{p}_k \cdot (\mathbf{x}_k - \mathbf{x}_{k-1}) + \Omega \Delta t \sum_{j=1}^M \left[ e^{i \mathbf{p}_k \cdot \boldsymbol{\nu}_j} - 1 \right] \tilde{a}_j(\mathbf{x}_{k-1})} \ .
\end{split}
\end{equation}
Now add in $T-1$ factors of $\Omega^N/\Omega^N$:
\begin{equation} \label{eq:pathint_CME_V2}
\begin{split}
P = &\lim_{T \to \infty} \left[ \frac{1}{\Omega^N} \sum_{\mathbf{n}_1} \right] \cdots \left[ \frac{1}{\Omega^N} \sum_{\mathbf{n}_{T-1}} \right] \int  \left( \frac{\Omega}{2\pi} \right)^N d\mathbf{p}_1 \cdots \int \left( \frac{\Omega}{2\pi} \right)^N d\mathbf{p}_T \\
& \exp{\sum_{k = 1}^{T} - i \Omega \ \mathbf{p}_k \cdot (\mathbf{x}_k - \mathbf{x}_{k-1}) + \Omega \Delta t \sum_{j=1}^M \left[ e^{i \mathbf{p}_k \cdot \boldsymbol{\nu}_j} - 1 \right] \tilde{a}_j(\mathbf{x}_{k-1})} \ .
\end{split}
\end{equation}
Riemann sums will play the role that the Euler-Maclaurin formula did (i.e. converting sums to integrals) in our earlier derivation. Recall that the (right endpoint) Riemann sum for a function $f$ on $[0, b]$ reads \cite{rudin1976}
\begin{equation}
\int_0^b f(x) dx \approx \sum_{i = 0}^{N} f(i \Delta x) \Delta x
\end{equation}
where $\Delta x = b/N$. If we take $b \to \infty$ and $N \to \infty$ in such a way that $\Delta x$ remains constant, we can write
\begin{equation} \label{eq:riemann1D}
\int_0^{\infty} f(x) dx \approx \sum_{i = 0}^{\infty} f(i \Delta x) \Delta x \ .
\end{equation}
The corresponding $N$-dimensional result is
\begin{equation} \label{eq:riemannND}
\begin{split}
& \int_0^{\infty} dx_1 \cdots \int_0^{\infty} dx_N \ f(x) \\ \approx & \sum_{i_1 = 0}^{\infty} \cdots \sum_{i_N = 0}^{\infty} f(i_1 \Delta x, \cdots, i_N \Delta x) (\Delta x)^N \ .
\end{split}
\end{equation}
Since the inverse system volume $1/\Omega$ seems to play the role of $\Delta x$ in Eq. \ref{eq:pathint_CME_V2}, we can use this Riemann sum result to approximate each sum as
\begin{equation}
\frac{1}{\Omega^N} \sum_{\mathbf{n}} \approx \int_0^{\infty} dx_1 \cdots \int_0^{\infty} dx_N
\end{equation}
so that our path integral is now
\begin{equation} \label{eq:pathint_CME_V3}
\begin{split}
P = &\lim_{T \to \infty} \int d\mathbf{x}_1 \cdots \int d\mathbf{x}_{T-1} \int  \left( \frac{\Omega}{2\pi} \right)^N d\mathbf{p}_1 \cdots \int \left( \frac{\Omega}{2\pi} \right)^N d\mathbf{p}_T \\
& \exp{\sum_{k = 1}^{T} - i \Omega \ \mathbf{p}_k \cdot (\mathbf{x}_k - \mathbf{x}_{k-1}) + \Omega \Delta t \sum_{j=1}^M \left[ e^{i \mathbf{p}_k \cdot \boldsymbol{\nu}_j} - 1 \right] \tilde{a}_j(\mathbf{x}_{k-1})} \ .
\end{split}
\end{equation}
Now we can argue just as we did in Sec. \ref{sec:con2}. Because we are taking $\Omega$ to be extraordinarily large in the thermodynamic limit,
\begin{equation}
\begin{split}
& \Omega \left[ e^{i \mathbf{p}_k \cdot \boldsymbol{\nu}_j} - 1 \right] \tilde{a}_j(\mathbf{x}_{k-1}) \tau  \\
 \approx & \Omega \left[ i \mathbf{p}_k \cdot \boldsymbol{\nu}_j - \frac{1}{2} \sum_{\ell = 1}^N \sum_{\ell' = 1}^N  p_k^{\ell} p_k^{\ell'} \nu_{j \ell} \nu_{j \ell'}  \right] \tilde{a}_j(\mathbf{x}_{k-1}) \tau   
 \end{split}
\end{equation}
i.e. $\Omega$ is so large that above term does not change much in value when Taylor expanded to second order in $\mathbf{p}_k$. Finally, we have
\begin{equation} \label{eq:pathint_CME_Vfinal}
\begin{split}
P = &\lim_{T \to \infty} \int d\mathbf{x}_1 \cdots \int d\mathbf{x}_{T-1} \int  \left( \frac{\Omega}{2\pi} \right)^N d\mathbf{p}_1 \cdots \int \left( \frac{\Omega}{2\pi} \right)^N d\mathbf{p}_T \\
& \exp{- \Omega \sum_{k = 1}^{T} \left[ i  \mathbf{p}_k \cdot \left( \frac{\mathbf{x}_k - \mathbf{x}_{k-1}}{\Delta t} - \sum_{j=1}^M \boldsymbol{\nu}_j \tilde{a}_j(\mathbf{x}_{k-1}) \right) +  \frac{1}{2} \sum_{\ell = 1}^N \sum_{\ell' = 1}^N \sum_{j=1}^M p_k^{\ell} p_k^{\ell'} \nu_{j \ell} \nu_{j \ell'} \tilde{a}_j(\mathbf{x}_{k-1}) \right] \Delta t}
\end{split}
\end{equation}
which is the same as the result from Sec. \ref{sec:con2} (c.f. Eq. \ref{eq:MSRJD_restricted}) but with additional factors of $\Omega$. It also exactly matches the system volume MSRJD path integral for the Fokker-Planck equation (c.f. Eq. 94 in Sec. V of our earlier paper\cite{vastola2019}). In other words, we have indeed derived a path integral equivalent to a set of Langevin equations/a Fokker-Planck equation. Moreover, it is equivalent to the \textit{same} set of Langevin equations that Eq. \ref{eq:MSRJD_restricted} is (as is easily seen after changing back from concentration variables to the original number variables)---although the integration bounds on the path integral are different here.  

Given that this approach was significantly simpler (in both a technical and conceptual sense), why bother with Gillespie's derivation? There are a few good reasons.
\begin{itemize}
\item The approximations provided by Eq. \ref{eq:riemann1D} and Eq. \ref{eq:riemannND} are more mathematically dubious than the Euler-Maclaurin formula (Eq. \ref{eq:eulermaclaurin}), which is well-studied and has precisely expressed error bounds. 
\item The thermodynamic limit may not apply to most biochemical systems of interest, given that molecule numbers are often large but not overwhelmingly so, and that the system volume (for example, of a cell) is not large enough to prevent crowding \cite{ellis2001, minton2001, hofling2013, rivas2016} and boundary effects \cite{komarova2002, buceta2017, holmes2019} from being important. 
\item The system volume approach only \textit{applies} when our CME describes a well-stirred, dilute mix of chemicals held at fixed temperature in a very large box---but the CLE is known to be a useful approximate description of all sorts of other stochastic systems (e.g. spiking neurons, fluctuating population dynamics models, stock options). In these other situations, there is no clear notion of a control parameter analogous to $\Omega$.
\item The system volume approach misses the subtlety of the integration bounds associated with the chemical Langevin/chemical Fokker-Planck equations; as we pointed out at the end of the previous section, it is a nontrivial issue that the domain of the approximating CLE will generally not be $[0, \infty)^N$.
\end{itemize}

\section{Discussion}
\label{sec:discussion}

We constructed an original path integral description of the CME, and applied Gillespie's conditions (suitably interpreted) to it in order to derive a path integral known to be equivalent to the CLE. In some sense, the difference between the system size approach and Gillespie's approach to deriving the CLE is the difference between approximating sums as integrals via Riemann sums, and via the Euler-Maclaurin formula. As discussed at the end of the previous section, while both approximation techniques can be valid in the appropriate circumstances, the Euler-Maclaurin formula is more generally applicable and has better characterized correction terms. 

It is interesting to note that, although we began with an exact path integral that involved taking the limit $\Delta t \to 0$ (Eq. \ref{eq:pathint_CME}), we coarse-grained time to end up with a path integral with fixed time step $\tau$ that does not get taken to zero (Eq. \ref{eq:pathint_CME_final}). This leads to another sense in which the CLE is only an approximate description, since a true CLE/Fokker-Planck path integral (see our earlier paper\cite{vastola2019}) also involves taking the limit $\Delta t \to 0$. However, the idea of a `macroscopic' timescale was addressed by Gillespie himself in his original paper \cite{gillespie2000}. There, he offered an analogy to current in an electric circuit: we can freely write and manipulate the derivative $I := dq/dt$, and think about the limit $dt \to 0$, \textit{provided} we understand that we are not taking it to be \textit{so} small that shot noise effects start to matter. 

Because our argument applied to each reaction/propensity function separately, it can in principle be used construct path integrals for hybrid systems. In other words, just as Harris et al. \cite{harris2006, harris2009, iyengar2010} do, we can suppose that Gillespie's two conditions apply only to a subset of all reactions or species, and construct a path integral in which some species/reactions are treated CLE-style, while others are treated CME-style. Indeed, there should be a path integral way to view all of the hybrid constructions---based on molecule numbers or separations of time scales---referenced in Sec. \ref{sec:review}. These path integrals could then be used to extract large deviation results. 

It is unclear if Gillespie's conditions could be applied to the Doi-Peliti path integral \cite{doi1976, doi1976second, peliti1985, mattis1998} in order to recover a CLE-equivalent path integral. Part of the difficulty is that the Doi-Peliti construction involves integrating over coherent states, which contribute integrals over the whole real line in the expression for the propagator \cite{peliti1985}; it is not necessarily straightforward to associate these with sums or integrals over state space.

\section{Conclusion}
\label{sec:conclusion}

The chemical Langevin equation is usually derived using Gillespie's two conditions, or large system volume arguments; as we described, both methods have clear path integral analogues. Our results suggest that path integrals offer a useful and mathematically precise way of thinking about the relationship between different levels of approximation (e.g. CME and CLE), and about coarse-graining biochemical models more generally.


\begin{acknowledgments}
This work was supported by NSF Grant \# DMS 1562078.
\end{acknowledgments}

\bibliography{cle_bib, euler-mac, jchemphys_bib, other_bib, re_pathint}

\end{document}